\documentclass[journal]{IEEEtran}

\usepackage{times}
\usepackage{epsfig}
\usepackage{graphicx}
\usepackage{amsmath}
\usepackage{amssymb}
\usepackage{subcaption, xcolor, multirow}
\usepackage[pagebackref=true,breaklinks=true,letterpaper=true,colorlinks,bookmarks=false]{hyperref}

\def \ie{\emph{i.e.}}

\begin{document}

\title{Training Patch Analysis and Mining Skills for Image Restoration Deep Neural Networks}

\author{Jae~Woong~Soh,~\IEEEmembership{Student~Member,~IEEE,}
        and~Nam~Ik~Cho,~\IEEEmembership{Senior~Member,~IEEE}
        
\thanks{J. W. Soh and N. I. Cho are with the Department
of Electrical and Computer Engineering, Seoul National University, Seoul, Korea.
e-mail: nicho@snu.ac.kr}
\thanks{Manuscript received April 19, 2005; revised August 26, 2015.}}

\markboth{Journal of \LaTeX\ Class Files,~Vol.~14, No.~8, August~2015}%
{Shell \MakeLowercase{\textit{et al.}}: Bare Demo of IEEEtran.cls for IEEE Journals}

\maketitle

\begin{abstract}
There have been numerous image restoration methods based on deep convolutional neural networks (CNNs). However, most of the literature on this topic focused on the network architecture and loss functions, while less detailed on the training methods. Hence, some of the works are not easily reproducible because it is required to know the hidden training skills to obtain the same results. To be specific with the training dataset, few works discussed how to prepare and order the training image patches. Moreover, it requires a high cost to capture new datasets to train a restoration network for the real-world scene. Hence, we believe it is necessary to study the preparation and selection of training data. In this regard, we present an analysis of the training patches and explore the consequences of different patch extraction methods. Eventually, we propose a guideline for the patch extraction from given training images.
\end{abstract}


%
\IEEEpeerreviewmaketitle

\section{Introduction}
\IEEEPARstart{D}{eep} neural networks have shown dramatic breakthroughs in computer vision tasks, most of which were enabled by the availability of many labeled data. The deep networks are still considered uninterpretable black-boxes, but exhaustive experiments are revealing some of their behaviors. Many works have shown that the performance of deep networks highly depends on the quality of the training dataset, and some research showed the effect of the number of training samples \cite{ML1, ML2, GEN, HowMany}. With small-scale training data, deep networks tend to be \emph{overfitted} to the training data, which deteriorates the generalization performance. For this reason, ``how to select or gather the training data for supervised learning'' has been an important research area for increasing the accuracy and generalizability while using reasonable size data. Especially in high-level vision tasks, many previous works addressed this problem because the cost of the dataset and its annotation is quite expensive.

To cope with the data issue, researchers are actively studying semi-supervised or weakly-supervised learning \cite{Weak1, Weak2, Weak3}, and unsupervised learning methods \cite{Unsuper1, Unsuper2}. Also, unsupervised domain adaptation methods have been proposed to reduce the domain gap between large-scale synthetic paired data and unlabeled real-world data \cite{UDA1, UDA2, UDA3}. Recently, active learning \cite{Active} was also proposed, which is to select a certain amount of unlabeled data for the annotation based on the estimated uncertainty.

\begin{figure}[t]
	\begin{center}
		\centering
		\includegraphics[width=0.98\linewidth]{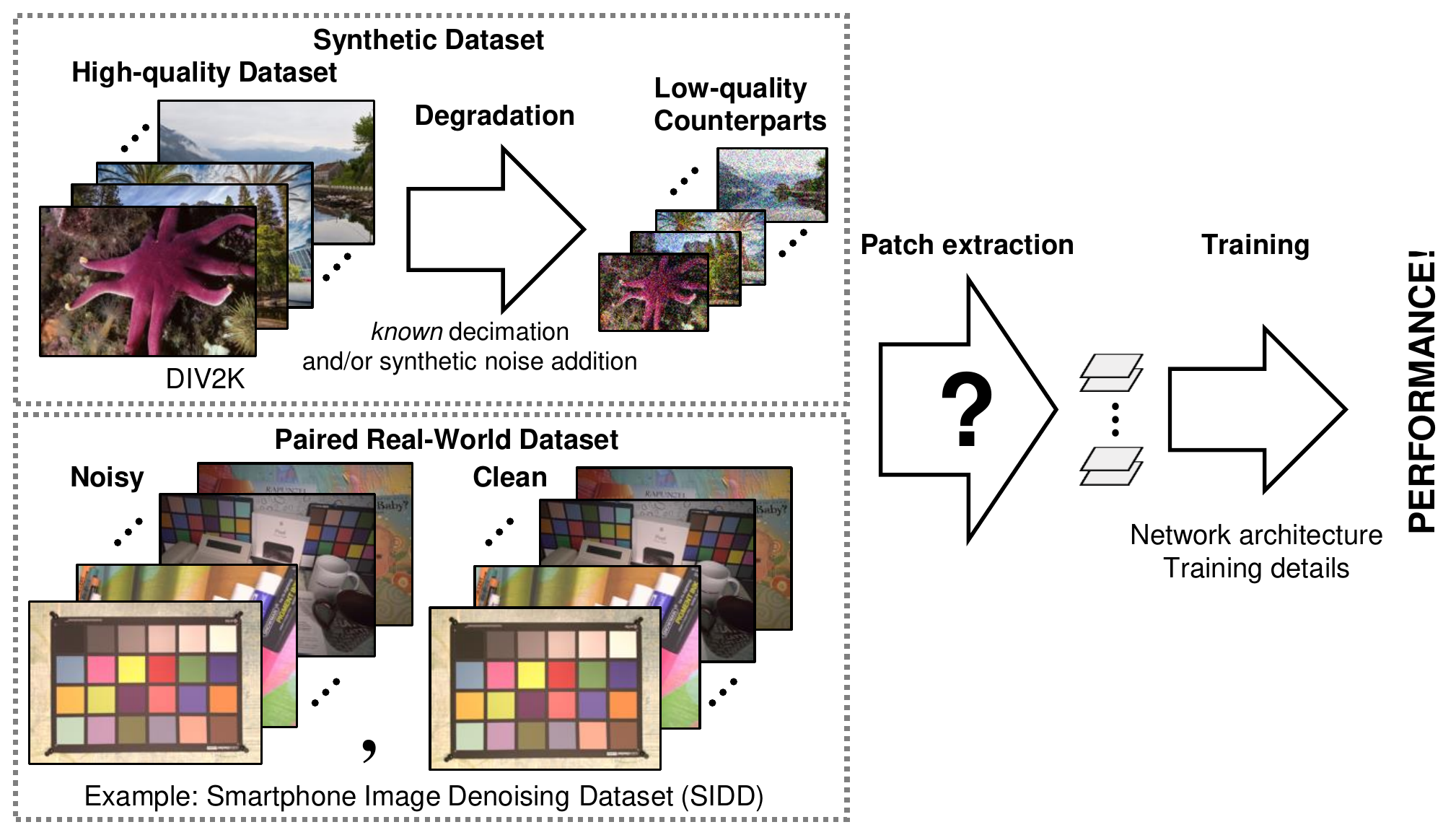}
	\end{center}
	\caption{For preparing synthetic data to train a restoration network, we can make an arbitrarily large number of paired data assuming that we know the image degradation model (upper dashed box). In this case, we can raise many questions about how to prepare the patches from the dataset. Recently, it is shown that the restoration networks overfitted to the synthetic data yield poor performance to the real-world degradations, and thus we need to (re)train the network with the real-world degraded data. In this case (lower box), the appropriate selection of effective patches is more important due to the deficiency of training data.}
	\label{fig:motivation}
\end{figure}

Meanwhile, deep neural networks are also actively used for low-level vision tasks such as image denoising \cite{DnCNN, FFDNet, ATDNet} and super-resolution \cite{Cyber3, CARN, RDN, ESRGAN, NatSR, SRFBN, OISR, Cyber2, Cyber4}. Notably, an important property of the image restoration problem is that it is mostly regarded as a ``self-supervised'' task when the degradation model is \emph{known}. In general, the observed image $\mathbf{y}$ is expressed as
\begin{equation}
\mathbf{y=Hx+n},
\label{imagemodel}
\end{equation}
where $\mathbf{x}$ is a latent clean image, $\mathbf{H}$ is a degradation matrix, and $\mathbf{n}$ is a noise. It means that if we have well-taken high-quality images, then we can degrade the high-quality images by using the \emph{known} degradation model to generate a large number of the paired dataset for supervised learning. Hence, since there seems to be no data deficiency problem in training the image restoration networks, recent methods mainly focused on the network architecture and loss function but less on the research of training datasets. In other words, the data-dependency of performance in low-level vision tasks was less frequently discussed. Even though several works discussed the training dataset, they were mainly the proposals of new datasets for image restoration in real-world scenarios \cite{Nam, Nah, RENOIR, SIDD, CameraSR}.

Regarding the above-stated problem, we raise several questions in this paper. We first raise obvious and intuitive simple questions:
\begin{enumerate}
\item Does overfitting occur in image restoration networks?
\item How good is the geometric (flip-rotation) data augmentation?
\end{enumerate}
Then, we further raise some controversial questions that may be informative and eventually can provide a guideline of patch mining skill:
\begin{enumerate}
\setcounter{enumi}{3}
\item What is the effect of the number of training patches?
\item How many patches are required to bring out the capacity of the network fully?
\item Is exploiting uncertain or hard samples effective for image restoration, as evidenced in high-level vision?
\end{enumerate}
Finally, we raise some questions regarding the generalization ability:
\begin{enumerate}
\setcounter{enumi}{6}
\item Are the good patches generic across different image restoration tasks?
\item Are the number of patches and the model size related?
\end{enumerate}

Our study is first motivated by unreproducible results of some restoration methods published in papers. Since the authors do not usually explain the details of their methods for preparing training patches, our reimplementations sometimes give inferior performances compared to the numbers in their original papers. Moreover, some recent works rely on the fortune (random crop) when extracting training patches \cite{DnCNN, FFDNet}. Some other works use a random selection of patches with empirical rules such as the gradient magnitudes or the variance of patches \cite{RDN, ARTN}. Interestingly, we found that even with the same dataset, each of the different patch extraction policies brings a noticeable performance gap. Furthermore,
data exploration is more important in preparing real-world training images because it is costly and difficult to capture well-registered pairs of clean-degraded images.

From the answers to the above questions and motivations, we finally present generic patch extraction and mining skills for image restoration. We show that our method improves the performance of some image restoration networks, and thus expect that it has the potential to boost the performance and convergence speed of general image restoration networks. We first analyze the effects of training data and then propose a guideline on extracting training patches for image restoration tasks.

Our motivation is summarized in \figurename{~\ref{fig:motivation}} and its caption, and contributions are summarized as:
\begin{itemize}
\item We attempt to find the answers to the above-stated questions, which have not been addressed well before.
\item We present a general guideline on patch extraction for image restoration.
\item We show that our strategy can be effectively applied to super-resolution and image denoising. Regarding the performance gain, we show that our patch mining skill is as useful as the recent neural architecture search (NAS)~\cite{FALSR,zophNAS} approach and the development of better architectures~\cite{OISR}.
\end{itemize}

\section{Related Work}

\subsection{Image Restoration based on Deep Networks}
Image denoising and super-resolution have been major areas of image restoration, and thus we address works related to these two areas. Typically, these two areas have long been studied \cite{BM3D, Cyber1, DnCNN, Cyber3}.

For the image denoising, DnCNN~\cite{DnCNN} adopts CNN with batch normalization \cite{BatchNorm} and rectified linear unit (ReLU), which shows great improvement compared to the conventional methods. FFDNet~\cite{FFDNet} is proposed to cope with multiple noise levels with a single CNN. Two-stage networks \cite{ATDNet, DUBD} are proposed for blind image denoising. Also, denoisers for real-world images have been proposed recently~\cite{CBDNet, Unprocess}.

In the case of CNN-based super-resolution, most of the works focused on the development of network architectures~\cite{Cyber3, VDSR, EDSR, CARN, DBPN, RDN, SRFBN, OISR, Cyber2, Cyber4}, while there are some works that studied loss function with generative models~\cite{SRGAN, SFT-GAN, ESRGAN, NatSR}, arbitrary scaling factor~\cite{MetaSR}, and unknown degradation model~\cite{BlindSR}.
There have been zero-shot approaches that rely on image-specific internal information instead of external patches \cite{ZSSR, MZSR}.
It has also been shown that the NAS approach or developing new efficient building blocks increase the performance of conventional networks, as addressed in FALSR~\cite{FALSR} and OISR~\cite{OISR}.

On the other hand, our research is focused on the training data analysis and guidelines for patch extraction, which also improves the performance of the conventional network as large as the FALSR and OISR. Just by the re-selection of the patches, we can obtain the performance gain comparable to or better than the NAS approach and new complex architectures.

\subsection{Data-relevant Strategies for Deep Networks}
There have been some observations that the performance of a network varies according to the property of training images. For example,
ESRGAN \cite{ESRGAN} addressed the effect of training patch size and demonstrated that the bigger the patch size, the better the performance. Also, there is some evidence that using a large number of high-quality training images brings better performance~\cite{DIV2K, EDSR, ESRGAN}. Moreover, a data augmentation method for image super-resolution has been proposed \cite{Cutblur}. However, in the case of real-world datasets for denoising~\cite{Nam, RENOIR, SIDD}, deblurring~\cite{Nah}, and super-resolution~\cite{CameraSR, RealSR}, which enabled more realistic image restorations than the synthetic data, the sizes of datasets are quite limited compared to synthetic cases. Thus, we attempt to derive a rule on extracting representative patches for training deep networks to bring out the best performance.

Other researches related to the dataset are hard example mining~\cite{FaceNet, Mining}, and active learning by learning loss~\cite{Active}. The hard example mining is a popular strategy for deep metric learning, which is shown to increase the performance and discriminative power of learned embeddings. One of our questions above is related to this problem, \ie, whether similar skills can be effective for image restoration. In the active learning research, they proposed to efficiently train the network when the budget for the annotation is limited. However, unlike high-level vision, it cannot be directly applied to real-scene image restoration because real-scene pairs need to be acquired simultaneously at the same spot. That is, it is impossible to generate a ground-truth image long after the low-quality image is taken.


\section{Backgrounds}
\subsection{Image Restoration}

\paragraph{Image Denoising}
Image denoising is to restore an original image with the assumption of $\mathbf{H=I}$ in Eq.~\ref{imagemodel}, \ie,
\begin{equation}
\mathbf{y=x+n}.
\end{equation}
The noise $\mathbf{n}$ is usually assumed to have a zero-mean white Gaussian distribution, but more realistic signal-dependent distributions are also considered.
However, all these assumptions do not perfectly match real-scene images. Hence, there are several works to train the network for the real-world noise recently \cite{CBDNet, Unprocess}.

\paragraph{Image Super-Resolution}
Image super-resolution is to obtain a plausible high-resolution image from a low-resolution input, where $\mathbf{H}$ is a composite operator of blur kernel and sub-sampling. It is expressed as
\begin{equation}
\mathbf{y=(x * k)\downarrow _{s} +n},
\end{equation}
where $\mathbf{k}$ is the blur kernel and $\mathbf{\downarrow_{s}}$ denotes decimation with scaling factor $\mathbf{s}$. Bicubic kernel and noise-free scenarios are mostly considered in the literature.

\section{Settings}
To analyze dataset dependencies in image restoration, we choose the single image super-resolution with the scaling factor of $\times 2$ as the main example.
We will also evaluate its dependencies for image denoising in later sections.

\subsection{Environments and Comparisons}
\subsubsection{Network Architecture}
Recently, many complicated architectures for the single-image super-resolution have been introduced, which exploit multi-kernels, dilated convolution, attention modules, etc. By employing such complex modules, researchers are reporting performance gains compared to the {\em EDSR baseline}\footnote{We consider {\em EDSR baseline} as the baseline in this paper because it has a simple architecture compared to the recent methods. Also, it is one of the popularly compared methods in the literature.} \cite{EDSR}. One of our validations is to show that our patch extraction method, when applied to the EDSR baseline, brings comparable or better performance than the recent complex models.

To be specific with the EDSR baseline, it consists of $16$ residual blocks where each convolution layer has $64$ output channels. The number of parameters of the baseline model is $1.37$ M. To increase the model capacity, we deepen the network by increasing the number of residual blocks in the baseline model.

\subsubsection{Training Dataset}
We use a high-quality dataset DIV2K \cite{DIV2K} for the training. MATLAB bicubic downsampling method is adopted to generate low-resolution images. We na\"{\i}vely extract $151,300$ patches from training images with the label patch size $96\times 96$ with stride $120$.

\subsubsection{Representative Values}
To find informative patches, we empirically define four metrics. They are chosen based on our hypothesis that a patch with rich texture and details is informative. The first metric is the uncertainty (we will call it ``loss'' shortly in the rest of this paper). We feed all training patches to a pretrained super-resolution network and measure the mean squared error (MSE) between the ground truth and output for each patch. Formally, the loss for the given patch $\mathbf{x}$ is defined as
\begin{equation}
Loss(\mathbf{x})=\frac{1}{HWC}||\mathbf{x}-f_\theta(\mathbf{y})||^2_2,
\label{eq:loss}
\end{equation}
where $f_\theta(\cdot)$ denotes the pretrained super-resolution network, and $\mathbf{y}$ is the degraded version of $\mathbf{x}$. Also, $H$, $W$, and $C$ denote height, width, and channel of the patch, respectively.
Second, we use the mean gradient magnitude value defined as 
\begin{equation}
Grad(\mathbf{x})=\frac{1}{HWC}\sum_{c}\sum_{h,w} ({|\nabla_h \mathbf{x}|^2 + |\nabla_w \mathbf{x}|^2}).
\end{equation}
Third, we consider the patch statistics, specifically we choose the standard deviation (std) of a patch.
Lastly, we consider high-frequency power (we will call it ``freq'' shortly in the rest of this paper), since the degradation-restoration process mostly harms the high-frequency components. It is expressed as
\begin{equation}
Freq(\mathbf{x})=\frac{1}{HWC}\sum_{c}\sum_{\omega_i, \omega_j \geq \frac{\pi}{2}} |DFT(\mathbf{x})|^2,
\end{equation}
which means that we adopt 2D DFT, especially the magnitude responses above $\frac{\pi}{2}$, as the measure of high-frequency power.

\subsection{Patch Statistics}

\begin{figure}[t]
\begin{center}
	\begin{subfigure}[t]{0.48\linewidth}
		\centering
		\includegraphics[width=1\columnwidth]{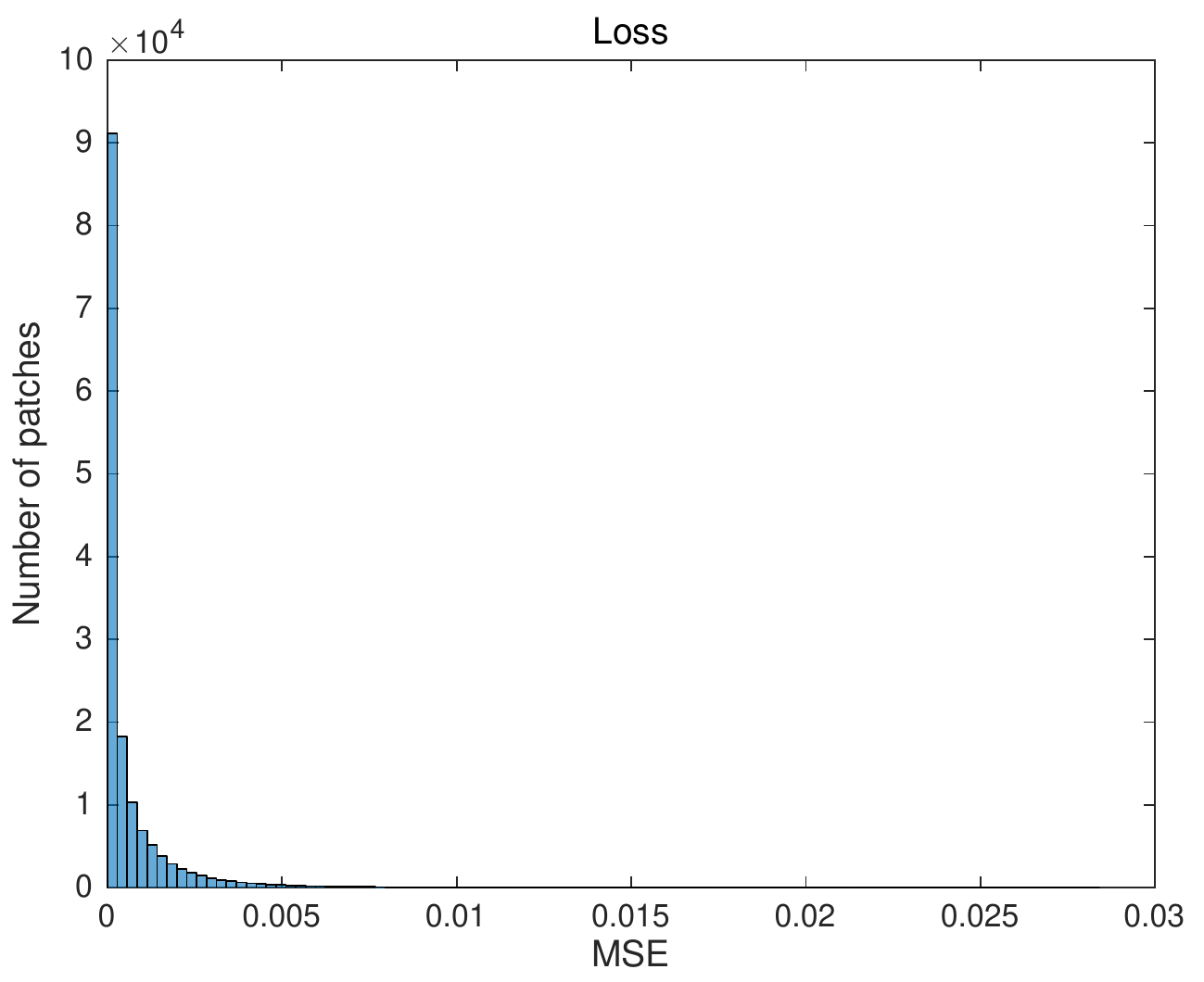}
		\caption{MSE of network prediction}
	\end{subfigure}
	\begin{subfigure}[t]{0.48\linewidth}
		\centering
		\includegraphics[width=1\columnwidth]{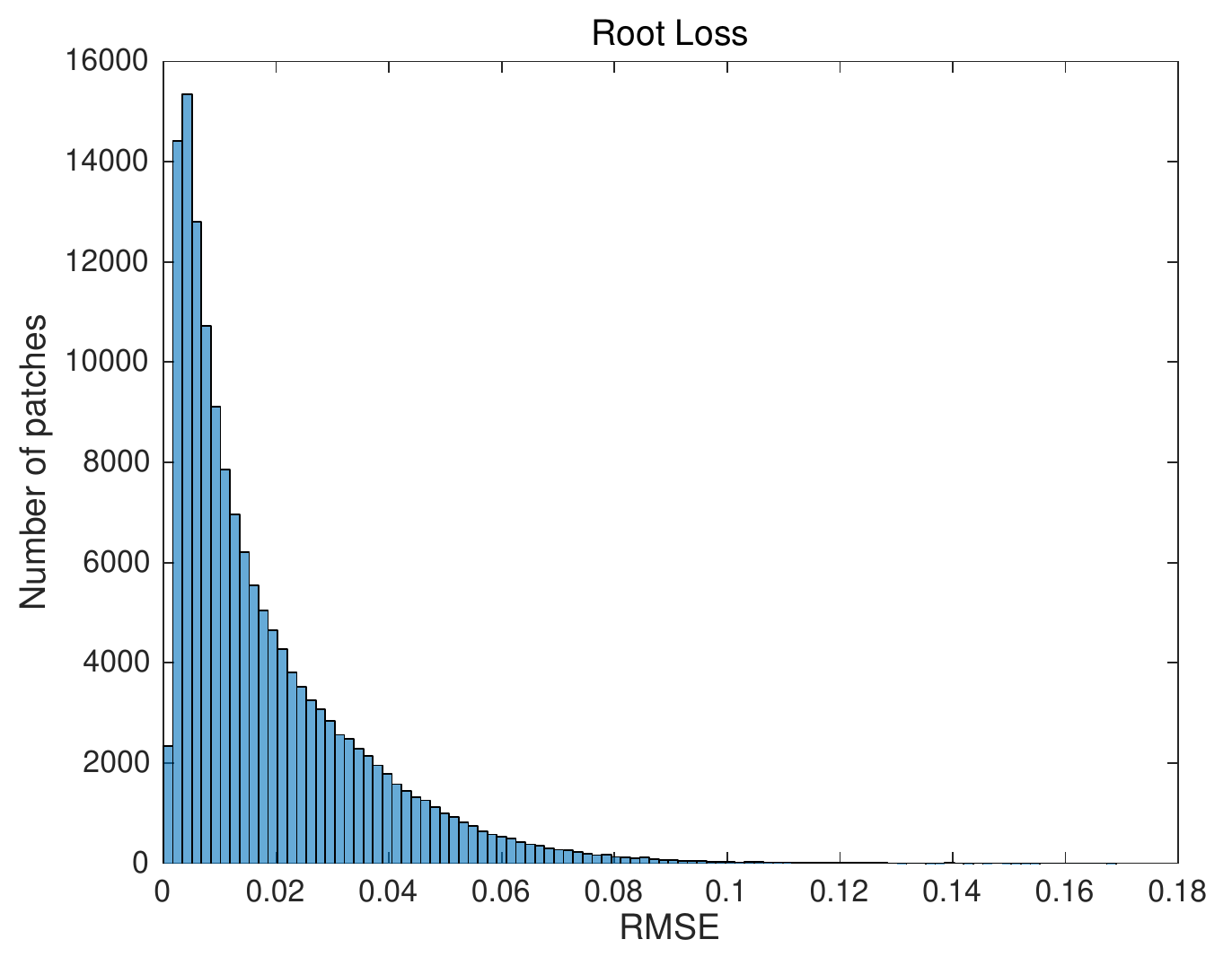}
		\caption{Root MSE}
	\end{subfigure}
\end{center}
\caption{Histogram of loss and square root of loss.}
\label{fig:hist_loss}
\end{figure}

\begin{figure}[t]
\begin{center}
	
	\begin{subfigure}[t]{0.48\linewidth}
		\centering
		\includegraphics[width=1\columnwidth]{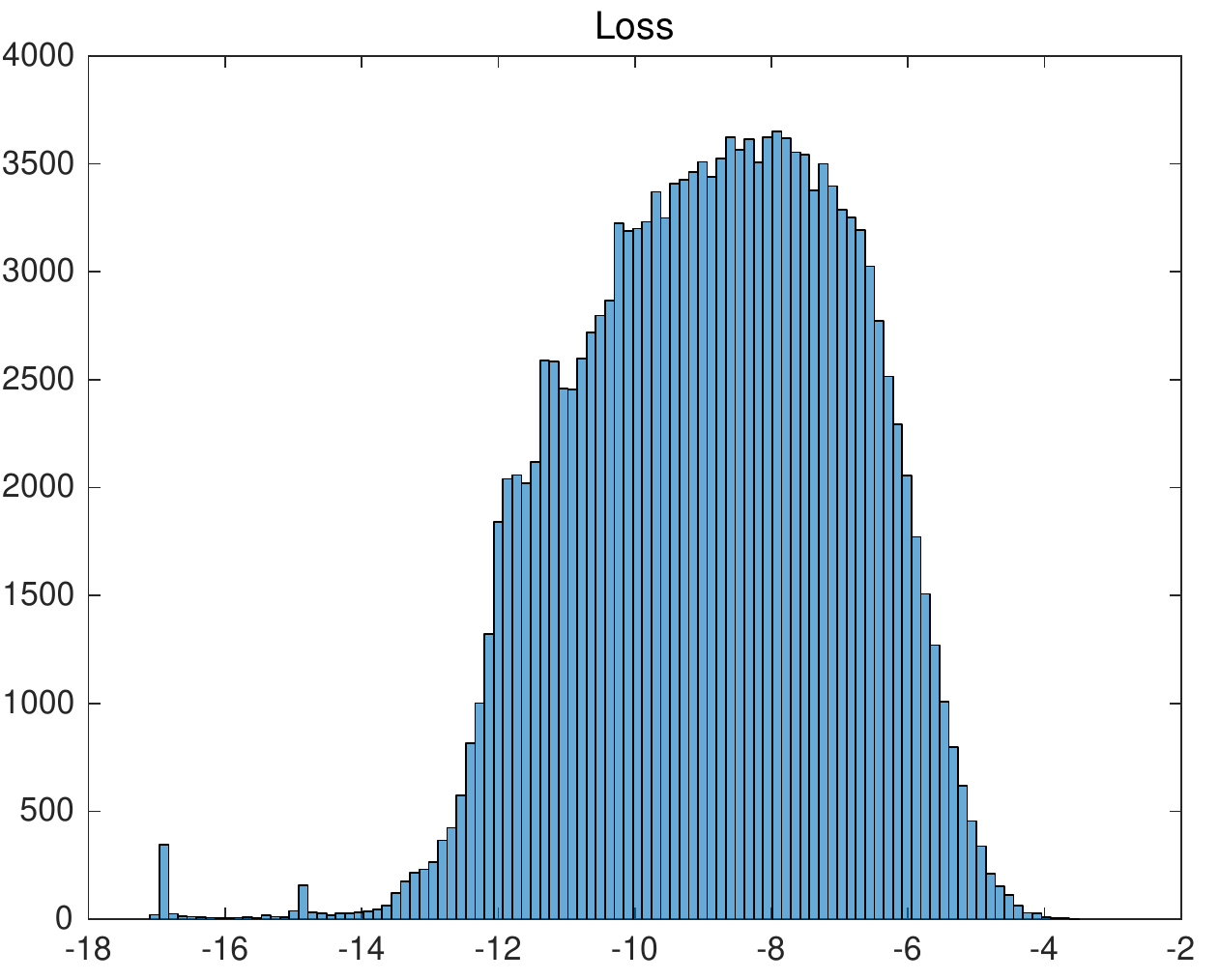}
		\caption{Uncertainty (Log-scale)}
	\end{subfigure}
	\begin{subfigure}[t]{0.48\linewidth}
		\centering
		\includegraphics[width=1\columnwidth]{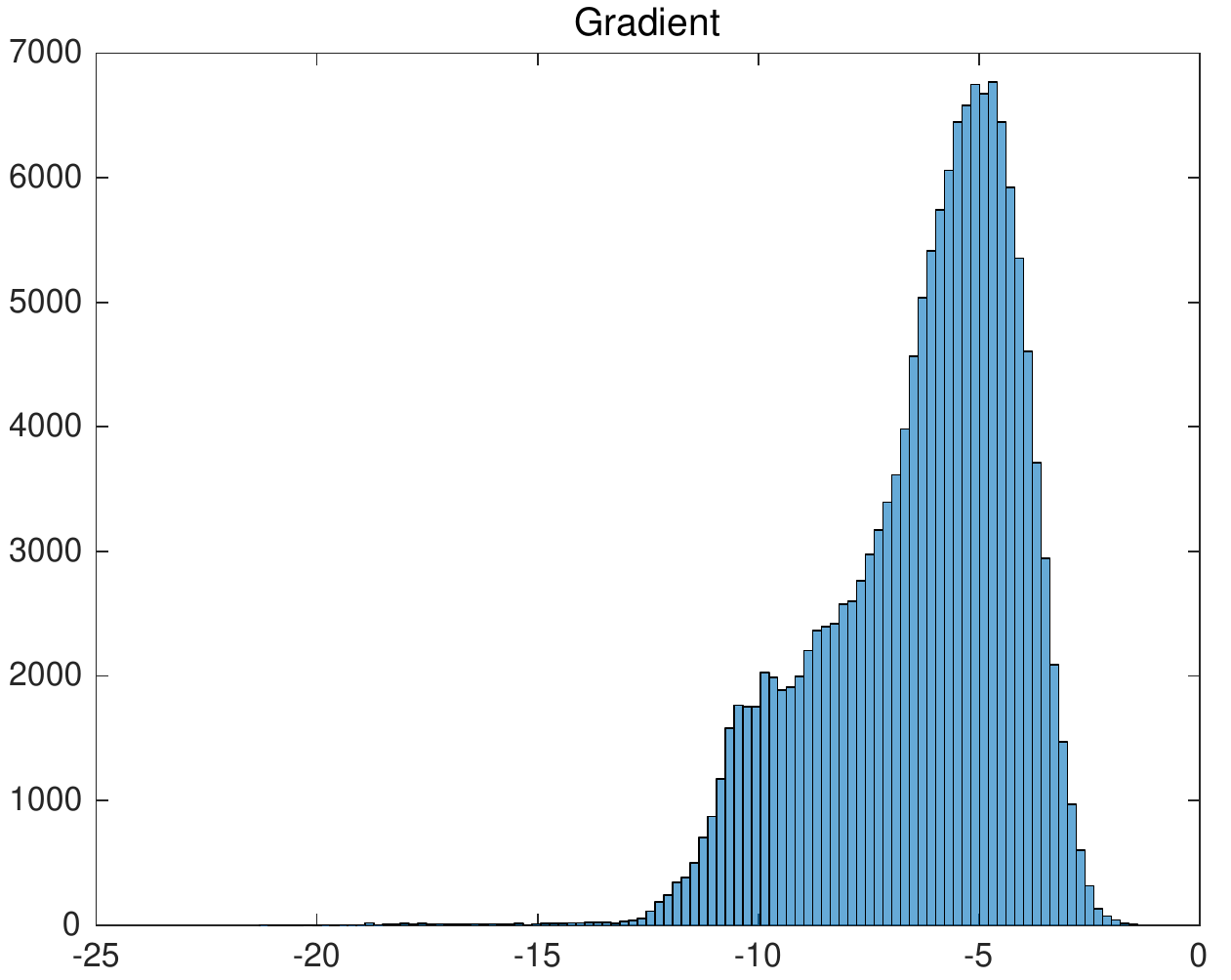}
		\caption{Gradient magnitude (Log-scale)}
	\end{subfigure}
	
	\begin{subfigure}[t]{0.48\linewidth}
		\centering
		\includegraphics[width=1\columnwidth]{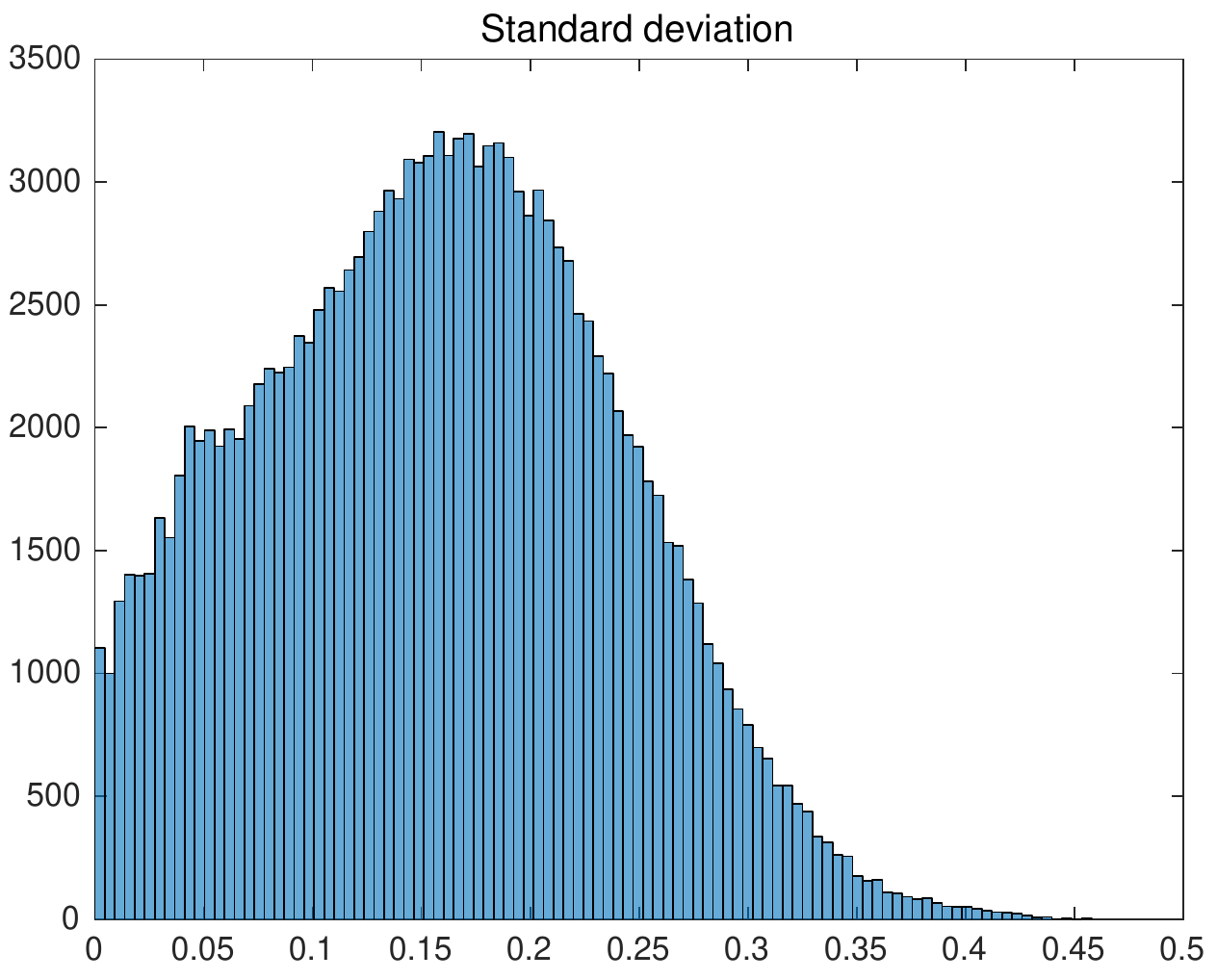}
		\caption{Standard deviation}
	\end{subfigure}
	\begin{subfigure}[t]{0.48\linewidth}
		\centering
		\includegraphics[width=1\columnwidth]{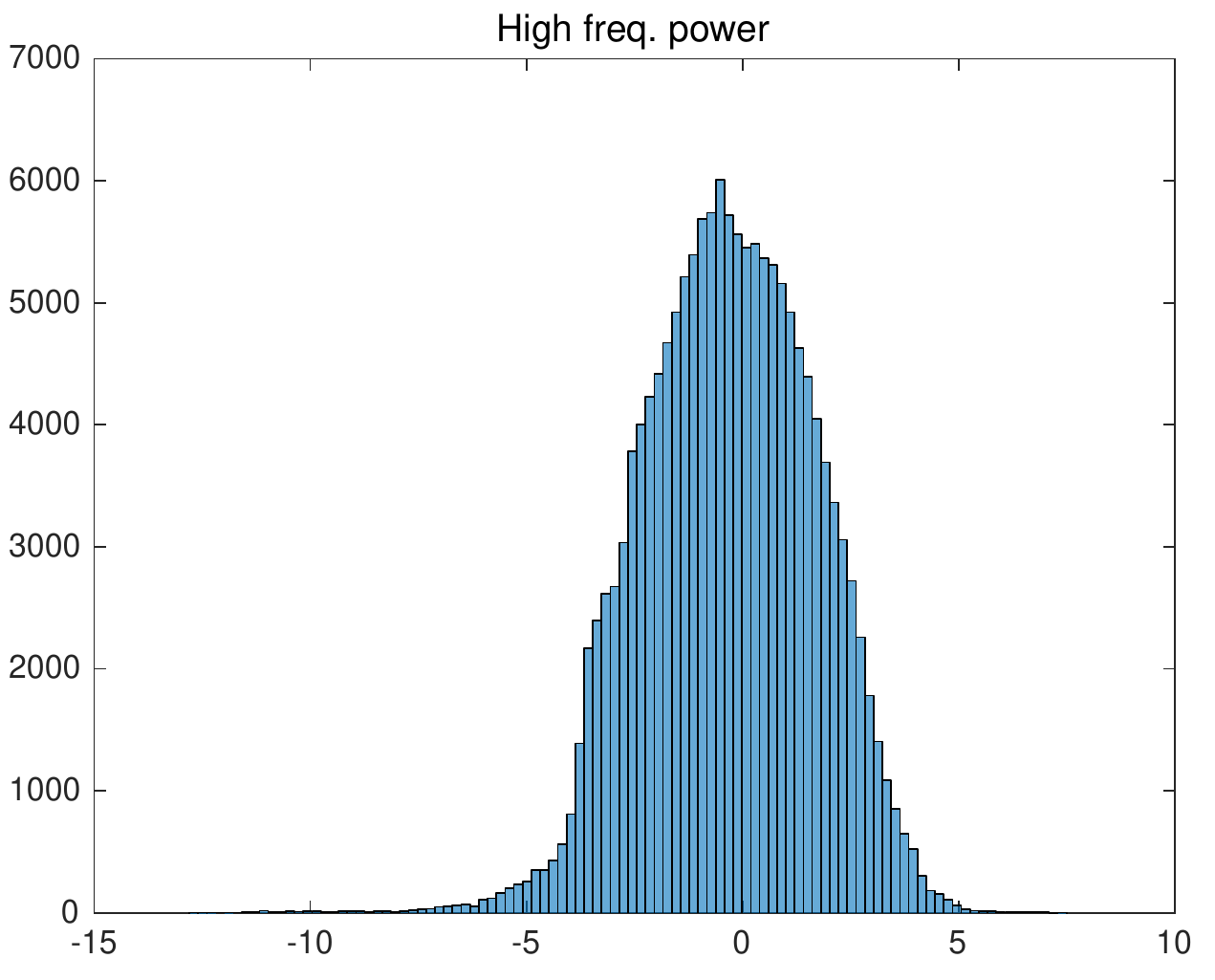}
		\caption{High freq. power (Log-scale)}
	\end{subfigure}
	
\end{center}
\caption{Histogram of the statistics of patches.}
\label{fig:hist}
\end{figure}

prediction (loss) is shown in \figurename{~\ref{fig:hist_loss}}.
It can be seen that the na\"{\i}ve extraction method brings too many patches with small MSE, \ie, redundant low-informative patches. We hypothesize that the na\"{\i}ve or random extraction of patches might be one of the reasons that slow down the convergence or make the network biased to the low-informative patches. Eventually, the network would hardly infer the areas such as high texture regions or vulnerable to aliasing artifacts (as the most performance loss by super-resolution CNN comes from blurred textures or aliased patterns). Also, other statistics of extracted patches by the metrics defined above are shown in \figurename{~\ref{fig:hist}}.

\section{Experiments}

For all experiments, we evaluate with Urban100 \cite{Self-Exemplar} set, which contains lots of textures and multi-scale recurrent patterns. Based on most research results, it has been shown that Urban100 set well discriminates the performance of super-resolution. For training, we utilize from $10,000$ to $150,000$ patches from total patches based on each criterion.

\subsection{Evaluations}
\subsubsection{Does overfitting occur in image restoration networks?}
\begin{figure}[t]
\begin{center}
	\begin{subfigure}[t]{0.48\linewidth}
		\centering
		\includegraphics[width=1\columnwidth]{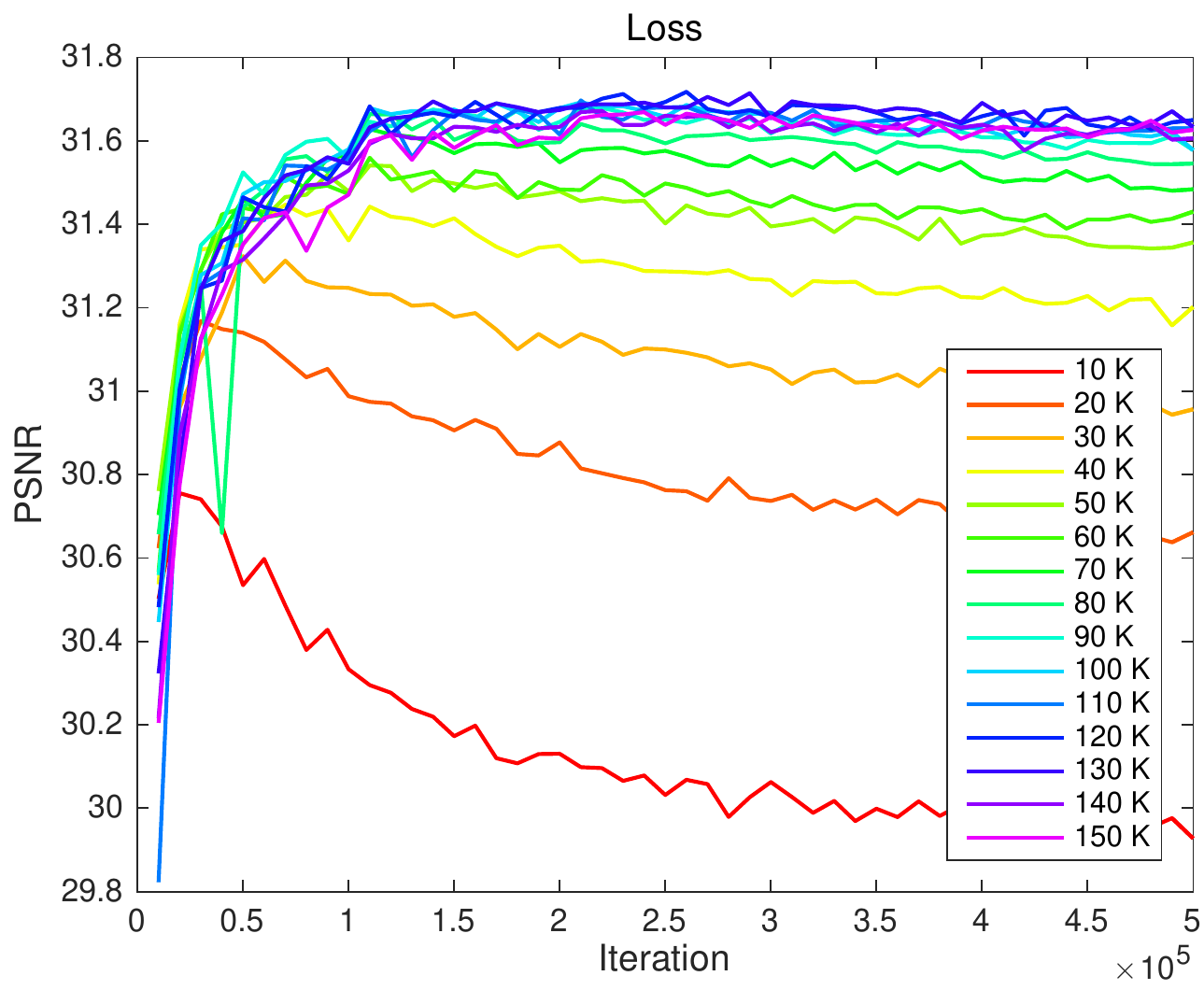}
		\caption{Without data augmentation}
		\label{fig:iter_reg}
	\end{subfigure}
	\begin{subfigure}[t]{0.48\linewidth}
		\centering
		\includegraphics[width=1\columnwidth]{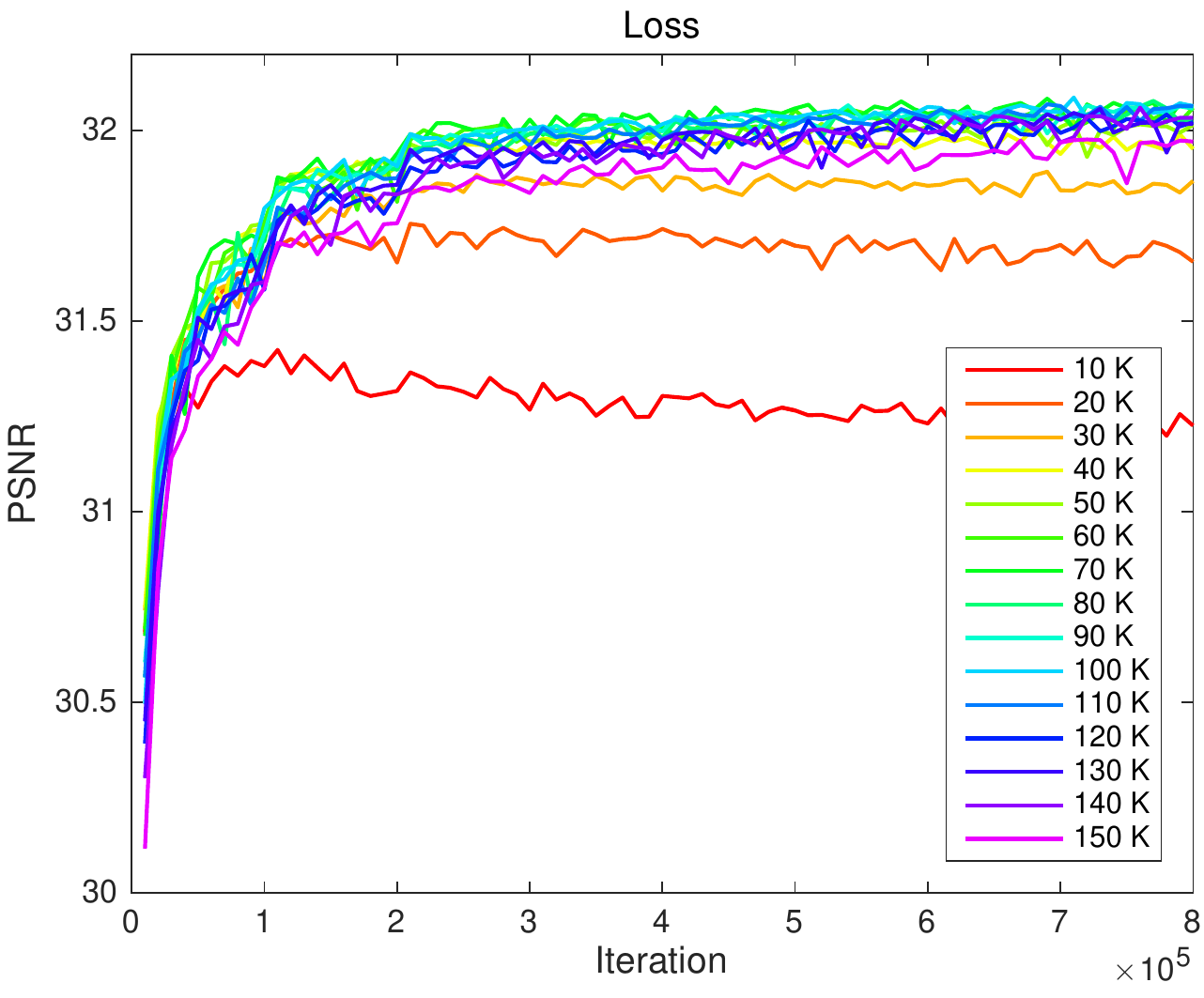}
		\caption{With data augmentation}
		\label{fig:iter_aug}
	\end{subfigure}
\end{center}
\caption{Convergence curve of the network across the number of training patches. ``Loss'' criterion is exploited when extracting patches.}
\label{fig:iter}
\end{figure}

It is evident and intuitive that overfitting occurs with small numbers of samples. However, our goal is more specific and manifest analysis; therefore, this paragraph provides a body of knowledge that may further help to think of overfitting problems.

\figurename{~\ref{fig:iter}} shows the convergence curves against the number of training patches.
To plot the convergence curves, we train the network based on the loss criteria with different numbers of patches.
It can be seen that more overfitting occurs with a fewer number of patches. As shown in \figurename{~\ref{fig:iter_reg}}, without data augmentation, most of results show severe overfitting where PSNR performance decreases after near $100,000$ iterations. Even with all training patches, subtle overfitting problems can be found. Flip and rotation augmentation skills can be a way to overcome such an overfitting problem, as shown in \figurename{~\ref{fig:iter_aug}}.
It is observed that at least $30,000$ patches are required to prevent overfitting when using geometric data augmentation.
The same tendency is found with other representative values.
In conclusion, an overfitting problem exists in image restoration tasks, and geometric data augmentation can be a solution to overcome overfitting. Precisely, at least $30$ K patches are required to prevent overfitting for the EDSR baseline.

\subsubsection{How good is the geometric data augmentation?}
\begin{figure}[t]
	\begin{center}
		\centering
		\includegraphics[width=0.8\columnwidth]{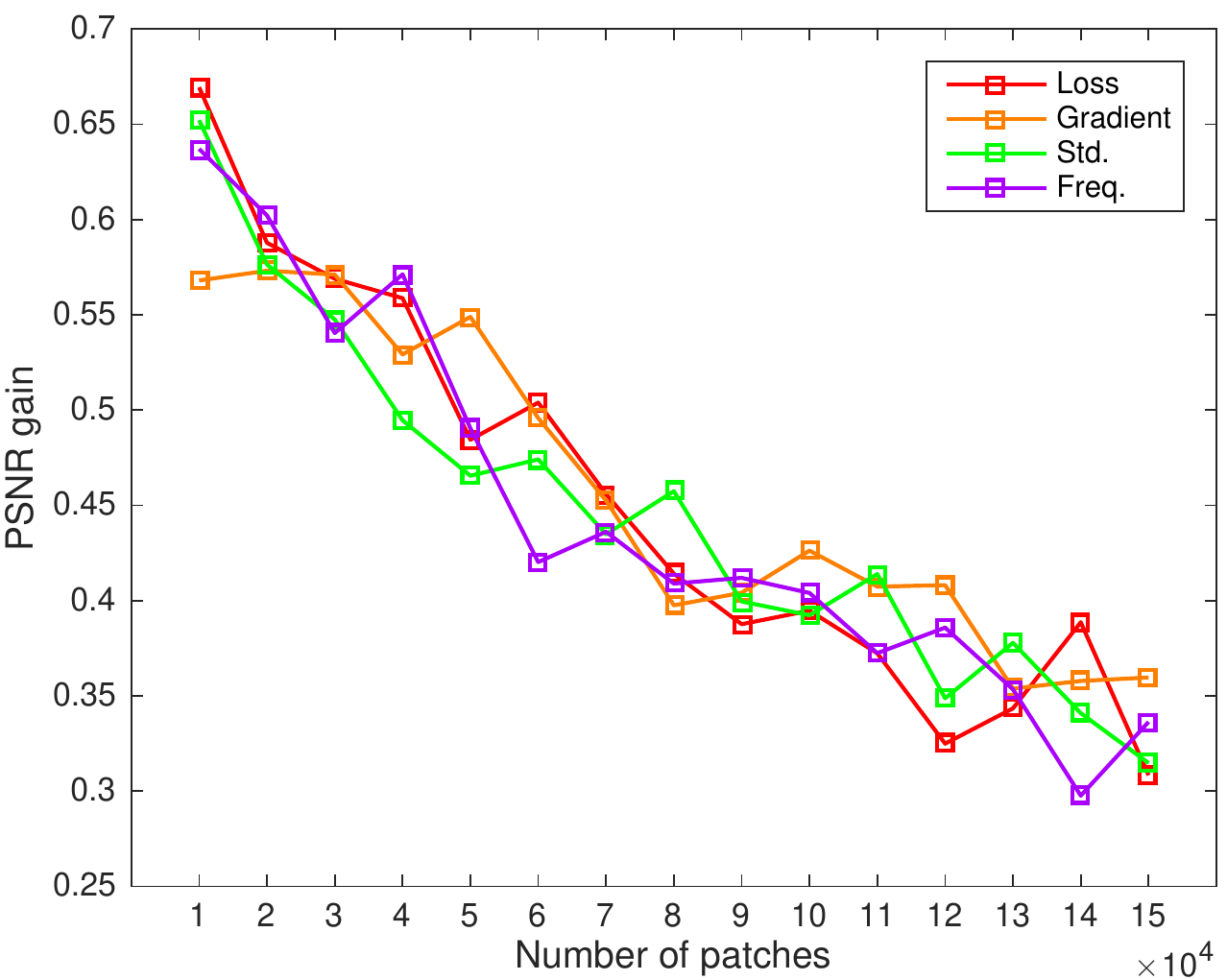}
	\end{center}
	\caption{PSNR gain due to geometric augmentation.}
	\label{fig:aug_effect}
\end{figure}

Data augmentation by exploiting flip-rotation invariance is a widely adopted skill for some computer vision tasks, including image restoration.
Notably, the use of geometric augmentation should be carefully considered depending on the restoration task where it should be invariant to such operations, \emph{e.g.}, spatially symmetric $\mathbf{H}$ and pixel-wise independent noise.
In \figurename{~\ref{fig:aug_effect}}, we explore how good the data augmentation is, with respect to the number of patches. It is found that the data augmentation is always good, boosting $0.3$ dB to $0.7$ dB PSNR gain. As shown, when the training patches are scarce, the effect of augmentation is quite big and begins to decrease when the number of patches increases. In conclusion, we may expect $0.5$ dB PSNR gain on average by using the data augmentation skill.

\subsubsection{What is the effect of the number of training patches?}
\begin{figure}[t]	
	\begin{center}	
		\begin{subfigure}[t]{0.48\linewidth}	
			\centering	
			\includegraphics[width=1\columnwidth]{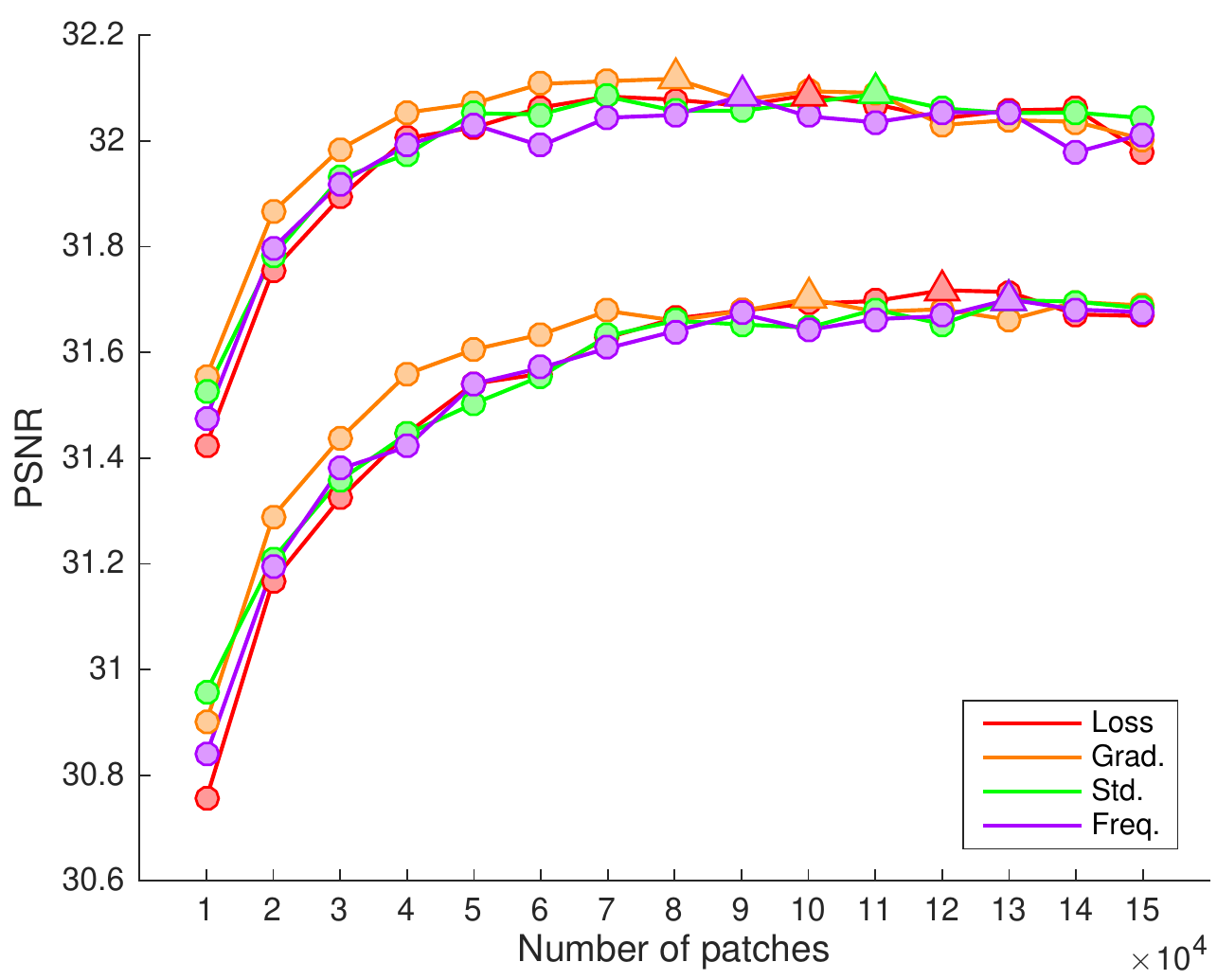}	
			\caption{All range of the number of patches.}	
			\label{fig:number_all}	
		\end{subfigure}	
		\begin{subfigure}[t]{0.48\linewidth}	
			\centering	
			\includegraphics[width=1\columnwidth]{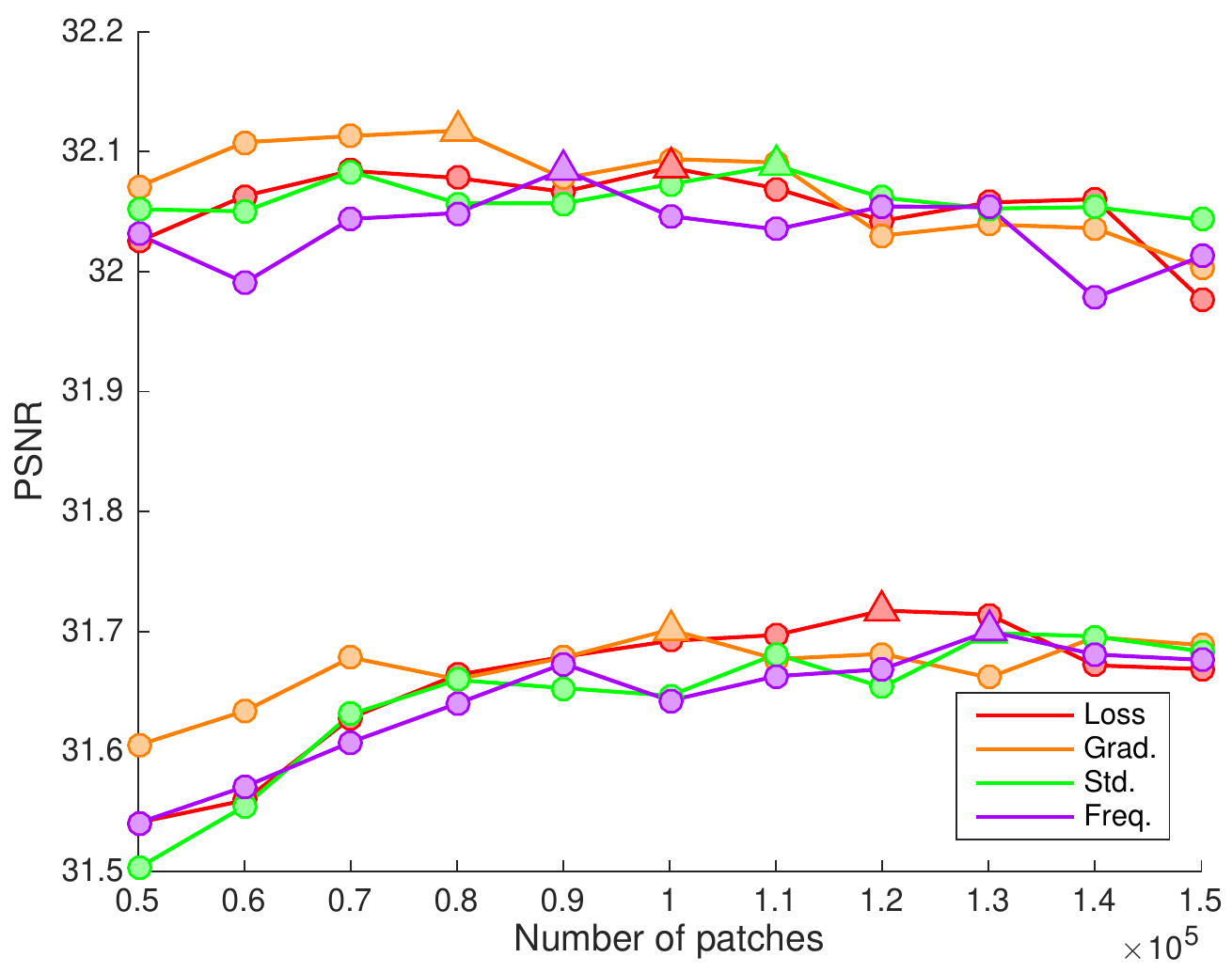}	
			\caption{Zoomed above $60$ K}	
			\label{fig:number_zoom}	
		\end{subfigure}	
	\end{center}	
	\caption{The number of patches vs. PSNR. We adopt early stopping to get the best result for each case. The best result for each case is highlighted with triangular markers. The upper curves are the results with data augmentation and the lower ones without data augmentation.}	
	\label{fig:number}	
\end{figure}

One may think that using ``as many training patches as possible'' might increase the performance of the network, but it may not be true in image restoration.
We visualized the PSNR results against the number of patches in \figurename{~\ref{fig:number}}.
Trivially, the performance tends to be increased with the increase in the number of training patches. However, as the data augmentation is adopted, which brings a similar effect of increasing the number of patches, using all the patches does not give the best performance across all criteria, as shown in \figurename{~\ref{fig:number}}. Interestingly, as the amount of patches exceeds a certain extent, the performance of the network drops. This is also observed in \figurename{~\ref{fig:iter}}, where we can see that most of the greenish curves are slightly above bluish curves (green colors use fewer patches than blue).

\subsubsection{How many patches are required to fully bring out the capacity of the network?}
From \figurename{~\ref{fig:number}}, without augmentation, the numbers above $130$ K show the best result while much smaller numbers ranging from $70$ K to $100$ K give the best with geometric data augmentation. It is recommended to exploit data augmentation skills and not use all of the total patches. Hence, when we have enough number of patches, it is suggested to use about half of them.

\begin{table}
	\caption{PSNR results against the number of training patches. `+' denotes the results with data augmentation.}
	\begin{center}
		\begin{tabular}{c|ccccc}
			\hline
			\rule[-1ex]{0pt}{3.5ex}
			\multirow{2}{*}{Criteria}& \multicolumn{5}{c}{Number of patches}  \\
			&$10$ K & $20$ K & $30$ K & $40$ K & $50$ K\\
			\hline\hline
			\rule[-1ex]{0pt}{3.5ex}
			Loss & 30.76 & 31.17 & 31.32 & 31.45 & 31.54\\
			Grad & 30.90 & 31.29 & 31.44 & 31.56 & 31.61\\
			\hline
			\rule[-1ex]{0pt}{3.5ex}
			Loss+ & 31.42&	31.76&	31.89&	32.01&	32.03\\
			Grad+ & 31.55&	31.87&	31.98&	32.05&	32.07 \\
			\hline
		\end{tabular}
	\end{center}
	\label{table:hard_sample}
\end{table}

\subsubsection{Is exploiting uncertain or hard samples good for image restoration as evidenced in high-level vision?}

One may think that using hard samples might be a good strategy in image restoration tasks. In other words, training with the collection of difficult patches, which make the ``loss'' in eq.~(\ref{eq:loss}), might be a desirable training method.
Interestingly, it is found that the restoration performance and hard examples are not related. In \tablename{~\ref{table:hard_sample}}, we explore the performance of network with small numbers of patches. ``Loss'' results correspond to hard sample mining based on network prediction error. ``Grad'' results are based on mining with a simple value of the magnitude of the gradient. As shown, with or without data augmentation, ``Grad'' results show better PSNR performance about $0.1$ dB in all cases. Unlike high-level vision tasks such as classification, image restoration is a regression task that is not to find decision boundaries but to find a well-generalizable mapping function. In this respect, we think it is better to provide a wide range of paired data than hard samples.

\subsubsection{Are the good patches generic across different image restoration tasks?}
\begin{figure}[t]
	\begin{center}
		\begin{subfigure}[t]{0.49\linewidth}
			\centering
			\includegraphics[width=1\columnwidth]{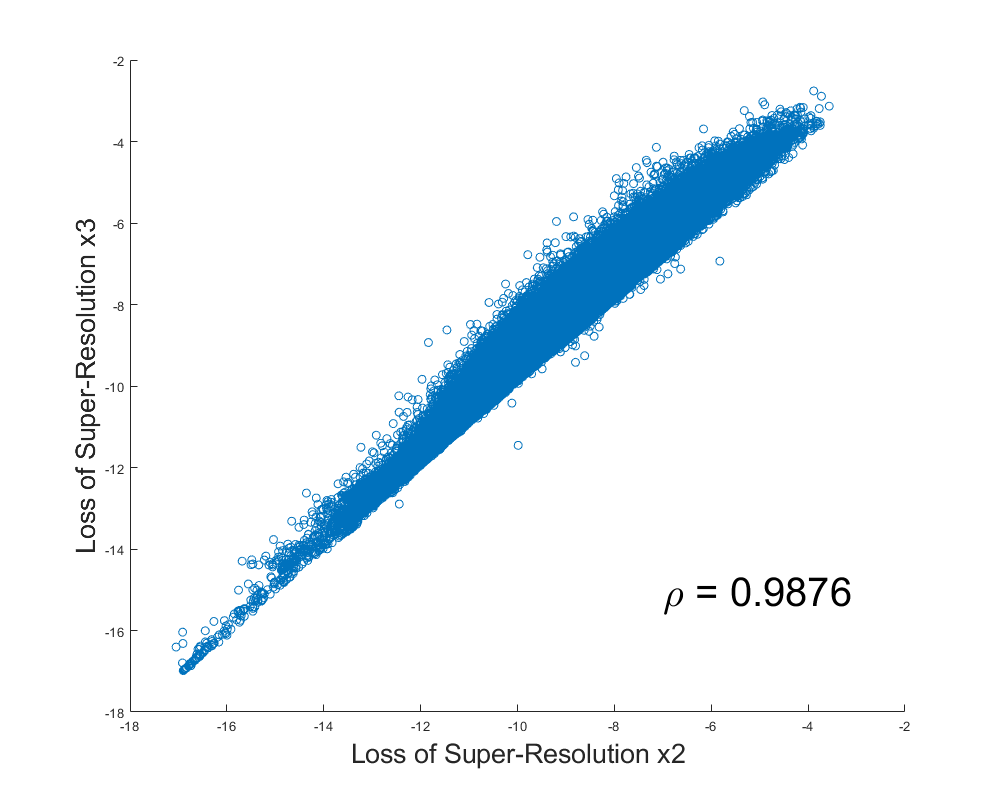}
			\caption{Loss of super-resolution with $\times 2$ vs. $\times 3$}
			\label{fig:x2_x3}
		\end{subfigure}
		\begin{subfigure}[t]{0.49\linewidth}
			\centering
			\includegraphics[width=1\columnwidth]{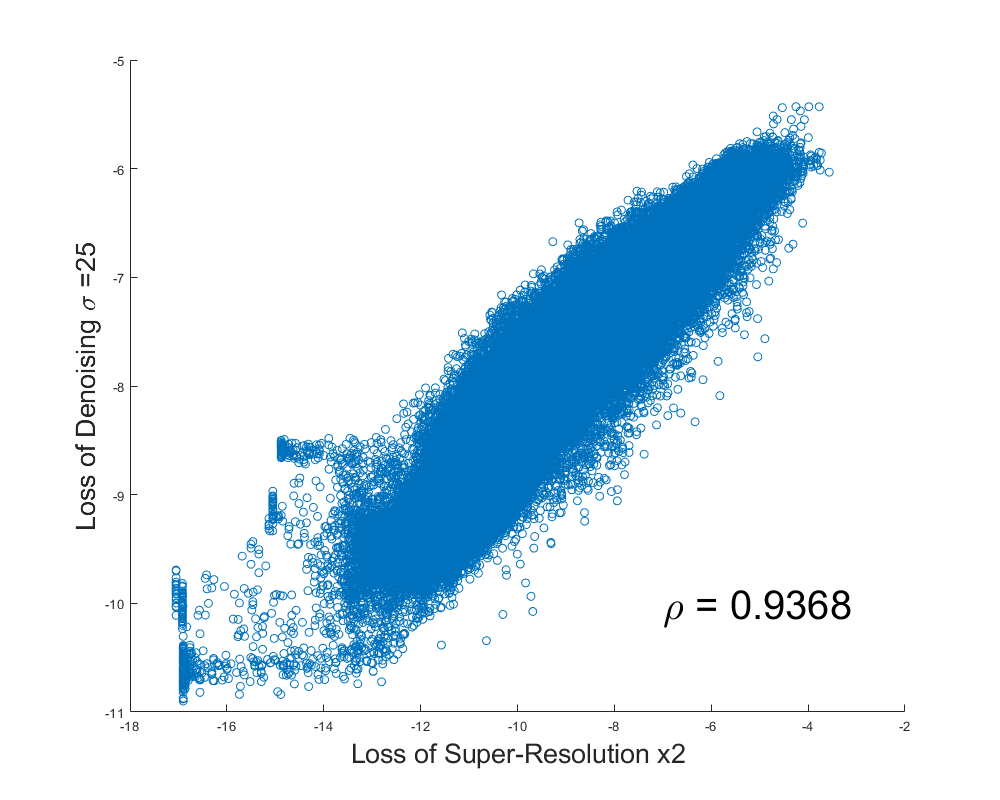}
			\caption{Loss of super-resolution with $\times 2$ vs. denoising $\sigma= 25$.}
			\label{fig:x2_dn}
		\end{subfigure}
	\end{center}
	\caption{Correlation among different image restoration tasks. We show the loss from a pretrained network of each task across all training patches. $\rho$ denotes the correlation coefficients between two tasks.}
	\label{fig:corr}
\end{figure}

To find a simple and generic guideline for patch extraction, we try to find the correlation of different image restoration tasks. Specifically, we adopt DnCNN \cite{DnCNN} with a noise level of $25$ as a representative denoiser. With the pretrained DnCNN, we calculate the MSE of prediction of each network, and the result is plotted in \figurename{~\ref{fig:corr}}. First, \figurename{~\ref{fig:x2_x3} shows the correlation within the task of super-resolution. It can be seen that super-resolution tasks of different scaling factors are highly correlated with the correlation coefficient of $0.9876$. Second, \figurename{~\ref{fig:x2_dn}} shows the correlation across the tasks: super-resolution and denoising. Although the correlation is lower than the former case, the two tasks are quite highly correlated with the correlation coefficient of $0.9368$.
	
From the graphs, we may conclude that the tendency of all experiments in our paper is generic across other image restoration tasks. For further verification, we also conduct experiments on the Gaussian image denoising task, which will be presented in Section ~\ref{sec:denoising}.

\subsubsection{Are the number of patches and the model size correlated?}
\begin{figure}[t]
	\begin{center}
		\centering
		\includegraphics[width=0.8\columnwidth]{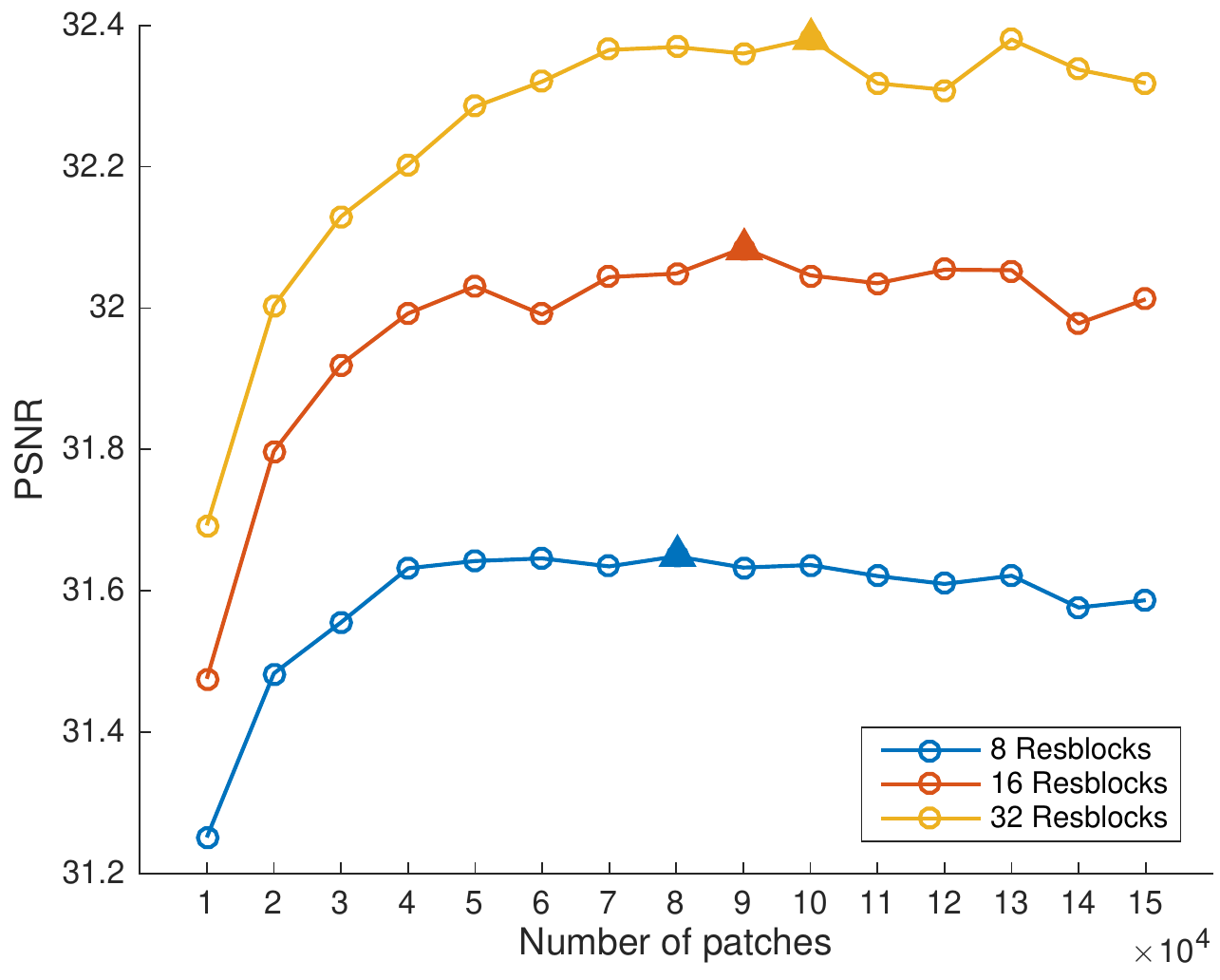}
	\end{center}
	\caption{The number of patches vs. PSNR with respect to the model size. Data augmentation is adopted for all three models. The best performances are highlighted with triangular markers.}
	\label{fig:model_size}
\end{figure}
From a recent research \cite{HowMany}, the number of samples needed for CNN has been discussed analytically, but it cannot be generalized to very deep high-dimensional non-linear CNNs. Thus, we need at least empirical researches for the CNN.
In this respect, to investigate and verify the generalization ability of patch extraction skills, we analyze the dependencies of model size and the number of training patches as shown in \figurename{~\ref{fig:model_size}}. We altered the number of residual blocks in the EDSR baseline to increase or decrease the model size. It can be observed that the behaviors are quite similar among the three models.

In summary, the training patches share similar capabilities and statistics regardless of the model size and tasks. Also, the number of training patches is a very important factor, similar to the network architecture. From the above-stated results, we expect that such a guideline of patch extraction would give similar consequences. Thus a good patch extraction skill might be generic to different image restoration tasks.

\begin{table}[!t]
	\caption{PSNR results with respect to the number of patches with different strides. The best results are highlighted with \textbf{bold}.}
	\begin{center}
		\begin{tabular}{|c|c|c|c|}
			\hline
			\rule[-1ex]{0pt}{3.5ex}
			Strides& 70 K & 80 K & 90 K\\
			\hline\hline
			\rule[-1ex]{0pt}{3.5ex}
			120 &\textbf{32.11} & \textbf{32.12} & 32.08\\
			60 & 32.10& 32.10 & \textbf{32.11}\\
			\hline
		\end{tabular}
	\end{center}
	\label{table:offset}
\end{table}

\begin{table}[!t]
	\caption{PSNR vs. number of patches for very large models.}
	\begin{center}
		\resizebox{0.7\linewidth}{!}{
			\begin{tabular}{|c|c|c|c|c|}
				\hline
				\rule[-1ex]{0pt}{3.5ex}
				Params & 60 K & 70 K & 80 K & 90 K\\
				\hline\hline
				\rule[-1ex]{0pt}{3.5ex}
				9.6 M & 32.74 & 32.73 & 32.67 & 32.68 \\
				\hline
			\end{tabular}
		}
	\end{center}
	\label{table:VeryLarge}
\end{table}

\subsection{Proposed Mining Skills} 

From the above discussions, we now go on our ultimate goal: ``how to extract good patches.'' We present a simple guideline for the good-patch extraction. As shown in \figurename{~\ref{fig:number}}, patch extraction based on \emph{gradient} metric outperforms other methods especially with the small number of patches. The gain is quite crucial, sometimes nearly $0.2$ dB.

What if we add more patches by reducing patch extraction strides? To assess this, we add about twice more patches by reducing strides from $120$ (non-overlap) to $60$ (overlap) and apply the same criteria. However, it does not give any gain as shown in \tablename{~\ref{table:offset}}, because additional patches are highly correlated to previously extracted ones. In other words, the patch size of $96$ and stride $120$ is enough to capture the whole DIV2K training dataset fully.

What about very big models? We also test on very large models by increasing the number of residual blocks to $128$. The results are shown in \tablename{~\ref{table:VeryLarge}}. Even with the large models, the results reach their peaks near half the number of total patches. We may say that the number of patches and the model size is less correlated, and our extracted patches can be adopted in any model. Additionally, data augmentation is always required, samples exceeding some extent of gradient are recommended, and at least near $80$ K number of patches are required.

In summary, our guideline is as follows.

\begin{enumerate}
	\item Collect images for the training dataset.
	\item If the number of the training dataset is enough, extract patches with the strides slightly bigger than the patch size (non-overlap). Otherwise, extract patches with overlap.
	\\
	- If we have less than $30$ K patches, we may not expect the performance due to overfitting.
	\item Discard half of patches with respect to their ``mean gradient magnitudes.''
	\item Augment the data using geometric (flip-rotation) transformation.
	\\
	- Consider flip-rotation invariance of the task.
\end{enumerate}

\section{Discussion}

\begin{table}
	\caption{The average PSNR/SSIM values on benchmarks.
		\emph{``Baseline'' results are from the official model trained by the authors.}}
	\begin{center}
		\resizebox{\linewidth}{!}{
			\begin{tabular}{|l|c|c|c|c|c|}
				\hline
				\rule[-1ex]{0pt}{3.5ex}
				$\times 2$ & Params & Set5 & Set14 & BSD100 & Urban100\\
				\hline\hline
				\rule[-1ex]{0pt}{3.5ex}
				Baseline \cite{EDSR} & 1.37 M & 37.90/0.9594 & 33.54/0.9167 & 32.14/0.8986 & 31.95/0.9264\\
				\hline
				\rule[-1ex]{0pt}{3.5ex}
				CARN \cite{CARN} & 1.59 M & 37.76/0.9590 & 33.52/0.9166 & 32.09/0.8978 & 31.92/0.9256\\
				\hline
				\rule[-1ex]{0pt}{3.5ex}
				FALSR-A \cite{FALSR} & 1.02 M & 37.82/\textcolor{black}{0.9595} & 33.55/0.9168 & 32.12/0.8987 & 31.93/0.9256\\
				\hline
				\rule[-1ex]{0pt}{3.5ex}
				OISR-RK2-s \cite{OISR} & 1.37 M & \textcolor{black}{37.98}/\textcolor{black}{0.9604} & \textcolor{black}{33.58}/\textcolor{black}{0.9172} & \textcolor{black}{32.18}/\textcolor{black}{0.8996}& \textcolor{black}{32.09}/\textcolor{black}{0.9281}\\
				\hline\hline
				\rule[-1ex]{0pt}{3.5ex}
				Ours (EDSR Baseline) & 1.37 M & \textcolor{black}{37.93}/0.9594 & \textcolor{black}{33.59}/\textcolor{black}{0.9170} & \textcolor{black}{32.17}/\textcolor{black}{0.8991}&\textcolor{black}{32.17}/\textcolor{black}{0.9284}\\
				\hline
				\rule[-1ex]{0pt}{3.5ex}
				Ours (OISR-RK2-s) & 1.37 M & 37.99/0.9596 & 33.69/0.9172 & 32.19/0.8993 & 32.26/0.9291 \\
				\hline
			\end{tabular}
		}
	\end{center}
	\label{table:PSNR}
\end{table}

\subsection{Comparisons with Benchmarks}

We compare our method with recent state-of-the-art super-resolution networks having similar complexity, on famous benchmark sets: Set5 \cite{Set5}, Set14 \cite{Set14}, BSD100 \cite{B100}, and Urban100 \cite{Self-Exemplar}, and list the results in \tablename{~\ref{table:PSNR}}. As stated previously, we adopt the relatively simple EDSR model as the baseline. The comparison shows that training the baseline with our patch selection strategy gives comparable or better performance than the recent complex networks. Our method achieves higher PSNR than the CARN \cite{CARN}, which is an efficient lightweight super-resolution model. Also, with simple residual blocks of EDSR, ours shows comparable results with FALSR~\cite{FALSR}, which is also an efficient lightweight structure optimized by NAS. We might say that the patch-search is as effective as the network-search in this case. OISR \cite{OISR} is a recent super-resolution model where the authors argued that it is possible to achieve a significant performance gain by replacing the building block of EDSR with their proposed one. Interestingly, our results are comparable to OISR, and we can also see that the patch selection is as effective as developing new architectures. With benchmark evaluation, it is found that selecting good training patches, rather than using the whole training dataset (all comparisons are trained with DIV2K \cite{DIV2K} training set), is an important factor that determines the performance.

Moreover, when we apply our dataset mining to the OISR-RK2-s \cite{OISR}, its performance on Urban100 \cite{Self-Exemplar} increases by about $0.2$ dB, which shows the generalization of our patch mining skill. We could not apply our method to the FALSR \cite{FALSR} because their code is not open.

\subsection{Results on Image Denoising}
\label{sec:denoising}

To further evaluate the effects of training patches on other image restoration tasks, we apply several training patch extraction settings to the Gaussian image denoising task. We experiment with a noise-level of $\sigma=25$, which has been considered a mid-level noise in former literature \cite{DnCNN, FFDNet}. We modified the baseline model by removing the upsampling module for super-resolution at the backend and adding the residual learning at the image level \cite{DnCNN}. We present results with different numbers of patches with three criteria: mean gradient magnitude (grad), standard deviation (std), and random extraction (rand). We evaluate RGB PSNR on Urban100 \cite{Self-Exemplar}. The results are shown in \figurename{~\ref{fig:denoising}}, where we can see that grad criterion shows the best results, and using about half of the patches also works here. We expect that our method can be applied to many other image restoration tasks.

\begin{figure}[t]
	\begin{center}
		\centering
		\includegraphics[width=0.8\columnwidth]{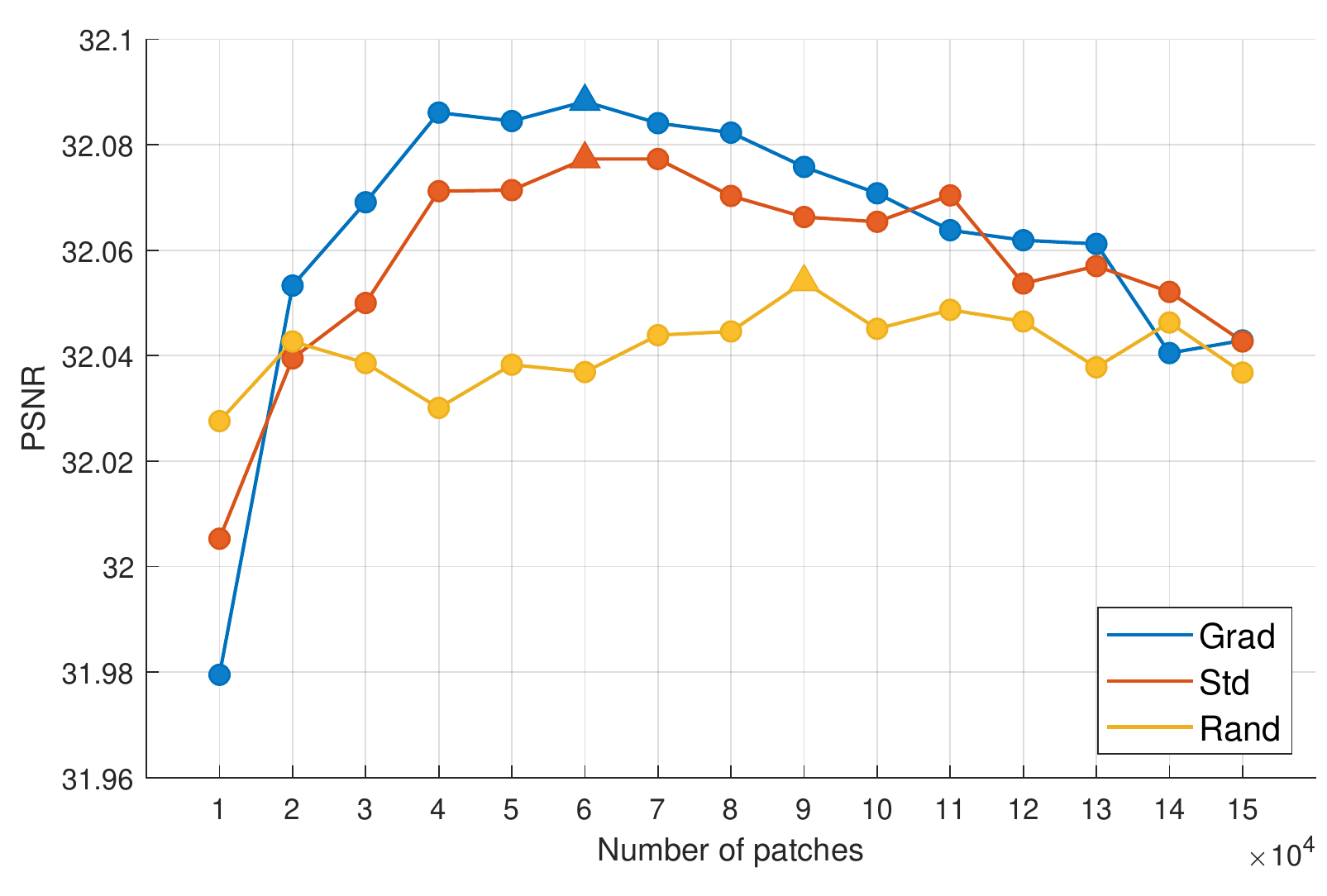}
	\end{center}
	\caption{Number of patches vs. PSNR with respect to different criteria. Data augmentation is adopted for all three models. The best performances are highlighted with triangular markers.}
	\label{fig:denoising}
\end{figure}

\section{Conclusion}
In this paper, we have considered patch extraction skills for CNN-based image restoration and presented simple rules to extract good patches. Our extensive experiments have shown the impact of different patch extraction skills, even from the same set of training images. Also, we demonstrated the effect of the number of patches. Eventually, we presented a simple patch extraction method and showed that it is as effective as the NAS-based method or new architecture development regarding the performance gain. Experiments and comparisons show that the proposed method is useful for some cases of single-image super-resolution and image denoising. We expect that the proposed method can also be applied to other image restoration problems.



\ifCLASSOPTIONcaptionsoff
  \newpage
\fi


%

%
%
%




\end{document}